\documentclass[prl,aps,amssymb,nofootinbib,twocolumn,showpacs]{revtex4}
\usepackage{amsmath}
\usepackage{amssymb}
\usepackage{amsthm}
\usepackage{amsfonts}
\usepackage{enumerate}
\usepackage{latexsym}
\usepackage[dvips]{graphicx}

\newcommand{\beq}{\begin{equation}}
\newcommand{\eneq}{\end{equation}}
\newcommand{\beqs}{\begin{equation*}}
\newcommand{\eneqs}{\end{equation*}}

\input{epsf}

\begin{document}

\tolerance 1000

\title{On Spin 1/2 Excitations in N\`eel ordered $2+ 1$ D
Antiferromagnets: Skyrmions in the $O(3)$ Nonlinear Sigma Model}

\author { Zaira Nazario$^\dagger$ and 
David I. Santiago$^{\dagger, \star}$ }

\affiliation{$\dagger$ Department of Physics, Stanford University,
             Stanford, California 94305 \\ 
	     $\star$ Gravity Probe B Relativity Mission, Stanford,
	     California 94305}
\begin{abstract}
\begin{center}

\parbox{14cm}{ We show that despite the absence of a Hopf term and zero Berry 
phase terms, the N\`eel ordered phase of $2+1$ D quantum antiferromagnets have
spin 1/2 excitations, i.e. {\it spinons}. The spinons are skyrmion 
excitations of a topological nature. Since skyrmion gap is proportional to
the spin stiffness, quantum criticality corresponds to skyrmion
gap collapse. We speculate that skyrmions are relevant at criticality and are, 
perhaps, related to recent suggestions of critical fractionalization.}

\end{center}
\end{abstract}
\date{\today}

\maketitle

It has been known for some time that $2+1$ dimensional short-range Heisenberg 
antiferromagnets in a bipartite lattice,
    \beq 
      \mathcal H = \sum_{<ij>} J \vec S_i \cdot \vec S_j \; ,
    \eneq 
with $J>0$ and the sum $<ij>$ is over nearest neighbor pairs, are described
by a relativistic field theory\cite{hal2,subir}: the $O(3)$ nonlinear sigma 
model augmented by Berry phases
    \begin{align}
      \begin{aligned}
	\mathcal Z =\int \mathcal D\vec n \delta(\vec n^2-1)
	e^{i\mathcal A} 
      \end{aligned}
    \end{align}
    \begin{align}
      \begin{aligned} \label{action}
	\mathcal A &= -S_B - \frac{1}{2}\int_{t_i}^{t_f} \! dt \!
	\int \!  d^2\vec x \rho_s \left[(\partial_{\vec x}\vec n)^2 -
	\frac{1}{c^2}(\partial_t\vec n)^2 \right] \,.
      \end{aligned}
    \end{align}
where  $\rho_s\equiv J S^2/\hbar$ is the spin stiffness, and
the spin-wave velocity $c=2 \sqrt{2} JSa/\hbar$ with 
$a$ being the lattice constant. The Berry phase terms are
        \begin{align}
	  \begin{aligned} \label{berry}
	    S_{B} &= \frac{S}{\hbar}\sum_i \epsilon_i \int_{t_i}^{t_f}
	    dt\int_0^1 du \, \vec n_i\cdot\left(\frac{\partial\vec
	    n_i}{\partial u}\times\frac{\partial\vec n_i}{\partial
	    t}\right) \,.
	  \end{aligned}
	\end{align}
We use a a square spatial lattice as regulator and we divide it in
sublattices A and B of even and odd sites respectively. $\epsilon_i$
is $1$ in sublattice A and $-1$ in sublattice B. The Berry phase terms
represent the sums of the areas swept by the vectors $\vec n_i(t,u)$
on the surface of a unit sphere as $u$ evolves it from $(0,0,1)$ to
$\vec n_i(t)$, and then $\vec n_i(t)$ evolves in time from its
direction at $t=t_i$ to the direction at $t=t_f$. This area
corresponds to the solid angle subtended by $\vec n$ on the
``so-called'' n-sphere it defines and does not depend on $u$ as it is
a geometrical invariant.

Forgetting the Berry phase terms for the moment, the $O(3)$ nonlinear
sigma model action (\ref{action}) has a classical ``ground state'' or
lowest energy state with N\`eel order. The ``ground state'' is a
solution of the Lagrange equations of motion for the nonlinear sigma
model corresponding to a constant vector which we chose to be $\vec
n=(0,0,-1)$. The action supports spin waves or Goldstone excitations
of arbitrary low energy which are time dependent solutions of the
linearized equations of motion. The equations of motion also have
static configurations, solitons, of finite energy
    \beq 
      E = \frac{\rho_s}{2} \int d^2\vec x (\partial_i\vec n)^2 =
      4\pi \hbar \rho_s \;.
    \eneq 
These solitons, first discovered by Skyrme\cite{skyrme}, are of a
topological nature as they are characterized by the integer winding
number
    \beq
      \label{q1} 
      q=\frac{1}{8 \pi} \int d^2 x \epsilon^{i j} \vec n \cdot \left(
      \partial_i \vec n \times \partial_j \vec n \right) \;.
    \eneq
These configurations consist in the order parameter rotating an integer number 
of times as one moves from infinity toward a fixed but arbitrary position in 
the plane. Since two dimensional space can be thought of as an infinite 2
dimensional sphere where the magnetic moments live, the excitations fall in 
homotopy classes of a 2D sphere into a 2D sphere: $S^2 \rightarrow S^2$. 
These topological configurations, skyrmions, are thus defined by the number of 
times they map the 2D sphere into itself as originally noticed by Belavin and 
Polyakov\cite{polya1}. Skyrmions are disordered at finite
length scales but relax into the N\'eel state far away: $
\lim_{|\vec x| \rightarrow \infty} \vec n=(0,0,-1)$. Skyrmions thus come 
with a directionality given by the direction of the N\`eel 
order they relax to at infinity. 

When one quantizes a classical theory, exact solutions of the
equations of motion correspond to eigenstates of the system. Goldstone
or spin-waves are thus the low-energy excitations of the N\`eel
ordered phase of the nonlinear sigma model.  Skyrmion configurations
are also solutions of the equations of motion. There are also
traveling skyrmions, i.e.  {\it time dependent} solutions
corresponding to skyrmions, obtained from the static skyrmion by
Lorentz transforming as the nonlinear sigma model is Lorentz
invariant. Thus skyrmions behave as excitations of the N\`eel ordered
phase and will disperse as solitons behave as quantum particle
eigenstates of field theories\cite{solitons}.

It has been known since the work of Wilczek and Zee\cite{wil}, WZ, that if we 
define the space-time current
    \beq 
      J^\mu= \frac{1}{8 \pi} \epsilon^{\mu \nu \sigma } \vec n \cdot
      \partial_\nu \vec n \times \partial_\sigma \vec n \, ,
    \eneq
it is conserved ($\partial_\mu J^\mu = 0$) and that the charge
associated with it is our topological charge: $q= \int d^2x J^0$.
Thus $q$ is a conserved quantum number. The conserved charge is called
the skyrmion number. Since $\partial_\mu J^\mu =0$, we can define a
gauge potential such that
    \beq
      J^\mu= \epsilon^{\mu \nu \sigma } \partial_\nu A_\sigma =
      \frac{1}{2}\epsilon^{\mu \nu \sigma } F_{\nu \sigma} \, .
    \eneq
WZ showed that if we modify the nonlinear sigma model action {\it
without} Berry phases by adding a term proportional to the Hopf
invariant
    \beq
      \Delta \mathcal A = -\frac{\theta}{2\pi}\int d^3x A_\mu J^\mu=
      -\frac{\theta}{4\pi} \int d^3x \epsilon^{\mu \nu \sigma } A_\mu
      F_{\nu \sigma} \, ,
    \eneq 
this increases the angular momentum of the skyrmion by
$\frac{\theta}{2 \pi}$ and skyrmions acquire an extra factor of
$(-1)^{\theta/\pi}$.

In the early days of high temperature superconductivity, this lead
several authors\cite{spinon} to propose that the effective low energy
physics of $2+1$ D Heisenberg antiferromagnets is a nonlinear sigma
model augmented by such a Hopf term with $\theta=\pi$. Thus it was
concluded that Heisenberg antiferromagnets have spin $1/2$ excitation,
i.e. spinons. According to WZ these spinons would be fermionic. Soon
afterwords it was shown\cite{hal2,nohopf} that the effective low
energy physics of $2+1$ D Heisenberg antiferromagnets is a nonlinear
sigma model augmented by Berry phase terms (\ref{berry}) but {\it no
Hopf term}. Moreover the Berry phase terms were shown to be
zero\cite{hal2,nohopf} if the N\`eel order parameter is continuous as
the contributions from each sublattice cancel. Thus such Berry phase
terms are important only in the disordered phase\cite{sachread}. In
the present work we show that despite the absence of Berry phase and
Hopf terms in the ordered phase of the $2+1$ D nonlinear sigma model,
the skyrmion configuration has spin 1/2. Therefore Heisenberg
antiferromagnets have spinon excitations which are gapped in their
ordered phase and these contribute to the physics of the system at
energy scales larger that the skyrmion gap, $4 \pi \hbar \rho_s$.

A very useful way of describing the $O(3)$ nonlinear sigma model is
through the stereographic projection\cite{polya1,gross1}:
    \beq
      \label{stereo}
      n^1 + i n^2 = \frac{2w}{|w|^2 + 1} \,
      n^3=\frac{1-|w|^2}{1+|w|^2} \, w=\frac{n^1 + i n^2}{1+n^3} \; .
    \eneq
In terms of $w$ the nonlinear $\sigma$-model Lagrangian is 
    \beq 
      \nonumber L = \frac{2}{g}\int d^2x \frac{\partial^\mu w
      \partial_\mu w^*}{ (1 + |w|^2)^2}=
    \eneq
    \beq 
      \label{nlsmlag}
      \frac{2}{g}\int d^2x \frac{\partial_0 w \partial_0 w^* - 2
      \partial_z w \partial_{z^*} w^* - 2 \partial_{z^*} w \partial_z
      w^*}{(1 + |w|^2)^2} \; ,
    \eneq
where $z = x + i y$ and $z^* = x -i y$ is its conjugate, and $1/g=\rho_s$. 
The classical equations of motion which follow by stationarity of the
classical action are 
    \begin{align}
      \Box w =& \frac{2 w^*}{1 + |w|^2}\partial^\mu w \partial_\mu w
      \text{ or}\\ \partial_0^2 w - 4\partial_z \partial_{z^*} w =&
      \frac{2 w^*}{1 + |w|^2} \left[ (\partial_0 w)^2 - 4 \partial_z w
      \partial_{z^*} w \right]
    \end{align}

The quantum mechanics of the $O(3)$ nonlinear sigma model is achieved
either via path integral or canonical quantization. The last is
performed by defining the momenta conjugate to $w$ and $w^*$, by
    \beq
      \Pi^*(t,\vec x) \equiv \frac{\delta L}{\delta \partial_0
      w(t,\vec x)} \, , \quad \Pi(t,\vec x) \equiv \frac{\delta
      L}{\delta \partial_0 w^*(t,\vec x)}\; ,
    \eneq
and then imposing canonical commutation relations among the
momenta and coordinates. The Hamiltonian is then given by 
    \begin{align}
      \begin{aligned}
	H &= \int d^2x \left( \Pi^* \cdot \partial_0 w + \Pi \cdot
	\partial_0 w^* - L\right) \\
	&= \int d^2x \left[ \frac{g}{2}(1 + |w|^2)^2 \Pi^* \Pi +
	\frac{4(\partial_z w \partial_{z^*} w^* + \partial_{z^*} w
	\partial_z w^*)}{g (1 + |w|^2)^2} \right]\; .
      \end{aligned}
    \end{align}
The Heisenberg equations of motion that follow from this Hamiltonian,
when properly ordered, are identical to the classical equations. 

We remind the reader that classically the lowest energy state is
N\`eel ordered for all $g < \infty$, i.e. the spin stiffness,
$\rho_s$, is never zero. Quantum mechanically the situation is
different. In $2+1$ D and higher dimensions, quantum mechanical
fluctuations cannot destroy the N\`eel order for the bare coupling
constant less than some critical value $g_c$\cite{polya2,halp}. At
$g_c$ the renormalized long-distance, low-energy coupling constant
diverges\cite{polya2,halp}, i.e. the system loses all spin
stiffness. At such a point quantum fluctuations destroy the N\`eel
order in the ground state as the renormalized stiffness vanishes. 

Linearization of the equations of motion leads to the low energy
excitations of the sigma model (magnons in the N\`eel phase and
triplons in the disordered phase) when quantized. We now turn our
attention to the N\`eel ordered phase. When the system N\`eel orders,
$\vec n$, or equivalently $w$, will acquire an expectation value:
    \beq
      \langle n^a \rangle = -\delta^{3a} \, , \quad \left \langle
      \frac{1}{w} \right \rangle = 0 \; .
    \eneq
where we have chosen the order parameter in the $-3-$direction as it
will always point in an arbitrary, but fixed direction. Small
fluctuations about the order parameter, $1/w = \nu$, are the magnon or
Goldstone excitations of the N\'eel phase. Its linearized equations of
motion are
    \beq
      \Box \nu=0 \, , \quad \partial_0^2 \nu - 4 \partial_z
      \partial_{z^*} \nu =0 \; .
    \eneq
Thus small amplitude excitations of the N\`eel phase have
relativistic dispersion that vanishes at long wavelengths as dictated
by Goldstone's theorem\cite{gold}. The magnons are of course spin $1$
particles. They have only 2 polarizations as they are transverse to
the N\`eel order.

The nonlinear sigma model possesses time independent skyrmion
solutions of a topological nature\cite{polya1,wil,gross1}. The
skyrmion number in terms of the stereographic variable, $w$, is
    \begin{align}
      \begin{aligned}
	q &=\frac{i}{2\pi} \int d^2 x \frac{\epsilon^{ij}\partial_i w
	\partial_j w^*}{ (1 + |w|^2)^2} \\
	= \frac{1}{\pi} & \int d^2 x \frac{\partial_z w \partial_{z^*}
	w^* - \partial_{z^*} w \partial_z w^*}{ (1 + |w|^2)^2}
	\label{q2} \; .
      \end{aligned}
    \end{align}
The skyrmion space-time current is given by
    \beq 
      J^\mu = \frac{i}{2\pi} \frac{\epsilon^{\mu \nu
      \sigma}\partial_\nu w \partial_\sigma w^*}{(1 + |w|^2)^2} \; .
    \eneq
From the expressions for the charge $q$ and for the Hamiltonian, it is
easily seen\cite{gross1,polya1} that $E \ge 4 \pi |q|/g$. We see
that we can construct skyrmions with $q > 0$ by imposing the condition
$\partial_{z^*} w =0$, that is $w$ is a function of $z$ only. Since the
magnetization, $\vec n$ or $w$, is a continuous function of $z$, the
worst singularities it can have are poles. The skyrmions will have a
location given by the positions of the poles or of the zeros of
$w$. Far away from its position, the field configuration will relax
back to the original N\`eel order. Therefore we have the boundary
condition $w(\infty)=\infty$, which implies
    \beq
      w=\frac{1}{\lambda^q}\prod_{i=1}^q(z-a_i) \, .
    \eneq

This configuration can easily be checked to have charge $q$ and energy
$4\pi q/g$. $\lambda^q$ is the arbitrary size and phase of the
configuration and $a_i$ are the positions of the skyrmions that
constitute the multiskyrmion configuration. The energy is independent
of the the size and phase due to the conformal invariance of the
configuration. We remark that since the multiskyrmions energy is the
sum of individual skyrmion energies, the skyrmions do not interact
among themselves\cite{gross1}. Similarly, the multiantiskyrmion
configuration can be shown to be
$w=\frac{1}{(\lambda^*)^q}\prod_{i=1}^q(z^*-a_i^*)$ with charge $-q$
and energy $4\pi q/g$.

We have just studied the skyrmion and antiskyrmion configurations
which relax to a N\`eel ordered configuration in the $-3$ direction
far away from their positions. We shall call them $-3$-skyrmions. The
skyrmion direction is given by the boundary conditions as
$z\rightarrow\infty$. For example, $(z-a)/\lambda$ gives $n^a(\infty)
= -\delta^{3a}$, so it is a $-3$-skyrmion. The $+3$-skyrmion is
$\lambda/(z-a)$. The $+1$-skyrmion is $(z-a)/(z-b)$. The $-1$-skyrmion
is $-(z-a)/(z-b)$. The $+2$-skyrmion is $i(z-a)/(z-b)$. The
$-2$-skyrmion is $-i(z-a)/(z-b)$. Because of the rotational invariance
of the underlying theory, they are all kinematically equivalent. They
are not dynamically equivalent since a N\`eel ordered ground state has
skyrmions and antiskyrmions corresponding to its ordering direction as
excitations.

We next map the $\pm 3$-skyrmions into $|+z\rangle$ and $|-z\rangle$
$SU(2)$ spins and show that we can define a superposition law by
multiplication of the configurations, such that they satisfy the spin
$1/2$ $SU(2)$ superposition law. We map
    \beq
      \frac{\lambda}{z-b} \Longleftrightarrow |+z\rangle \; , \qquad
      \frac{z-a}{\lambda} \Longleftrightarrow |-z\rangle
    \eneq
We map multiplication by a complex constant $\alpha$ of the $SU(2)$
spins into the skyrmions via $\lambda \rightarrow \lambda/\alpha$,
i.e. by changing the skyrmion size:
    \beq
      \alpha |+z\rangle \Longleftrightarrow
      \frac{\lambda}{\alpha(z-b)} \; , \qquad \alpha |-z\rangle
      \Longleftrightarrow \frac{\alpha(z-a)}{\lambda}
    \eneq
If we superpose the $+3$-skyrmion with the $-3$-skyrmion we
obtain
    \beq
      \frac{1}{\sqrt{2}}\left(|+z\rangle + |-z\rangle\right)
      \Longleftrightarrow \frac{z-a}{z-b}
    \eneq
The last is a $+1$-skyrmion, which maps into $|+x\rangle$. The
skyrmions obey the spin $1/2$ addition rule $(|+z\rangle +
|-z\rangle)/\sqrt{2} = |+x\rangle$.  If we superpose the $+3$-skyrmion
with the negative of the $-3$-skyrmion we obtain
    \beq
      \frac{1}{\sqrt{2}}\left(|+z\rangle - |-z\rangle\right)
      \Longleftrightarrow - \frac{z-a}{z-b}
    \eneq
The last is a $-1$-skyrmion, which maps into $|-x\rangle$. The
skyrmions obey the spin $1/2$ addition rule $(|+z\rangle -
|-z\rangle)/\sqrt{2} = |-x\rangle$. If we superpose the $+3$-skyrmion
with $i$ times the $-3$-skyrmion we obtain
    \beq
      \frac{1}{\sqrt{2}}\left(|+z\rangle + i|-z\rangle\right)
      \Longleftrightarrow i \frac{z-a}{z-b}
    \eneq
The last is a $+2$-skyrmion, which maps into $|+y\rangle$. The
skyrmions obey the spin $1/2$ addition rule $(|+z\rangle + i
|-z\rangle)/\sqrt{2} = |+y\rangle$. If we superpose the $+3$-skyrmion
with $-i$ times the $-3$-skyrmion we obtain
    \beq
      \frac{1}{\sqrt{2}}\left(|+z\rangle - i|-z\rangle\right)
      \Longleftrightarrow -i \frac{z-a}{z-b}
    \eneq
The last is a $-2$-skyrmion, which maps into $|-y\rangle$. The
skyrmions obey the spin $1/2$ rule addition $(|+z\rangle - i
|-z\rangle)/\sqrt{2} = |-y\rangle$.

The $O(3)$ invariance of the sigma model implies that superpositions
of the $+1$ and $-1$ skyrmions, and $+2$ and $-2$ skyrmions, also
satisfy the spin $1/2$, $SU(2)$ superposition law. This follows since
we could have chosen our stereographic projection for $w$ in terms of
$\vec n$ so that the skyrmions that look simple are the $1$ or
$2$-skyrmions instead of the $3$-skyrmions. Similarly we obtain that
antiskyrmions obey the spin $1/2$ $SU(2)$ superposition law. This
implies that skyrmions and antiskyrmions carry spin $1/2$. This is
surprising as the $2+1$ D nonlinear sigma model has no Berry phase
terms in its ordered phase and has no Hopf term if obtained from the
Heisenberg Hamiltonian as considered here.

In order to add weight to our conclusion the N\`eel phase of $2+1$D
Antiferromagnets has spin $1/2$ topological excitations we point out
that the spin of an object is determined by its rotational properties.
We thus take a look at the properties of skyrmions under rotations.
If we have a vector in the $+z$ direction and rotate it by an angle
$-\theta$ around the $+x$ direction we finish with a vector along the
direction $\sin\theta \hat y + \cos\theta \hat z$. The corresponding
rotation for half integral $SU(2)$ spins is given by
    \beq
      |\xi \rangle \equiv e^{\sigma_x\theta/2} \, |+z\rangle =
       \cos\left(\frac{\theta}{2}\right) |+z\rangle + i
       \sin\left(\frac{\theta}{2}\right) |-z\rangle
     \eneq
According to the map of the $SU(2)$ spins into skyrmions, we see that
we have to superpose
    \begin{align}
      \begin{aligned}
	\cos\left(\frac{\theta}{2}\right) |+z\rangle
	\Longleftrightarrow
	\frac{\lambda}{(z-b)\cos\left(\theta/2\right)} \\
	i \sin\left(\frac{\theta}{2}\right) |-z\rangle
	\Longleftrightarrow \frac{i
	\sin\left(\theta/2\right)(z-a)}{\lambda}
      \end{aligned}
    \end{align}
to obtain
    \beq
      |\xi \rangle \Longleftrightarrow
       i\tan\left(\frac{\theta}{2}\right) \frac{z-a}{z-b} \; .
     \eneq
The skyrmion direction is defined by its direction as
$z\rightarrow\infty$. We easily see that
$w(\infty)=i\tan(\theta/2)$. From the stereographic projection
(\ref{stereo}) we get that the skyrmion has direction $n_1 + i n_2 = i
\sin\theta$, $n_3 = \cos\theta$ or $n_i=\langle \xi | \sigma_i | \xi
\rangle$, exactly the rotational properties of a spin 1/2 object.

That skyrmions and antiskyrmions have spin 1/2 can be understood from
intuitive but correct reasoning. Since magnon excitations have zero
skyrmion number and skyrmion number is conserved, a magnon of nonzero
energy will have a nonzero matrix element for decay into equal number
of skyrmions and antiskyrmions. Consider a magnon with energy above
the threshold for decay into one skyrmion and one antiskyrmion, but
below the threshold for decay into larger numbers of skyrmions and
antiskyrmions. Since a magnon has spin one and the skyrmion and
antiskyrmion into which it decay can be in an S-wave state, the
skyrmion and antiskyrmion must have spin 1/2 by conservation of
angular momentum.

Therefore, skyrmions and antiskyrmions carry half integral angular
momentum, i.e. {\it they are spinons}. We see that the $2+1$ D $O(3)$
nonlinear sigma model in its ordered phase has excitations with spin
$1/2$ despite the absence of Hopf or Berry phase terms. This
conclusion is independent of whether the microscopic spins are
integral or half-integral.

Since the skyrmion gap is proportional to the spin stiffness which
vanishes at the quantum critical point where N\`eel order is lost,
spinon gap collapse corresponds to criticality. So for nearly
critical\cite{halp} but ordered $2+1$ D quantum antiferromagnets at
energies and temperatures larger than the skyrmion or spinon gap
(which will be quite small as it is proportional to the spin
stiffness), we expect the physics and thermodynamics to have important
contributions from the spinon fluctuations. This is perhaps the reason
why spinon phenomenology works for underdoped cuprates as they are
nearly critical and nearly $2 + 1$ D\cite{halp}. More importantly, it
has recently been proposed that there are new degrees of freedom at
quantum critical points\cite{bob1,sachdev2}, that these critical
excitations will be fractionalized, and that for $2 + 1$ D they will
be spinons. We suggest that the skyrmions or spinons of the ordered
phase are related to these critical spinons because they carry the
same quantum numbers and because the quantum critical point
corresponds to skyrmion gap collapse.


\begin{thebibliography}{99}

\bibitem{hal2} F. D. M. Haldane, Phys. Rev. Lett. {\bf 61}, 1029
  (1988)

\bibitem{subir} S. Sachdev, {\it Low Dimensional Quantum Field
Theories for Condensed Matter Physicists}, Proc. of the Trieste Summer
School 1992 (World Scientific, Singapore, 1994).

\bibitem{skyrme} T. Skyrme, Proc. Royal Soc. London A {\bf 260}, 127
  (1961).

\bibitem{polya1} A. A. Belavin and A. M. Polyakov, JETP Lett. {\bf
  22}, 245 (1975).

\bibitem{solitons} J. Goldstone and R. Jackiw, Phys. Rev. D {\bf 11},
  1486 (1975); J.-L. Gervais and B. Sakita, Phys. Rev. D {\bf 11},
  2943 (1975); N. H. Christ and T. D. Lee, Phys. Rev. D {\bf 12}, 1606
  (1975); E. Tomboulis, Phys. Rev. D {\bf 12}, 1678 (1975).

\bibitem{wil} F. Wilczek and A. Zee, Phys. Rev. Lett. {\bf 51}, 2250 (1983).
              
\bibitem{spinon} I. E. Dzyaloshinski, A. M. Polyakov and P. B. Wiegmann, Phys.
	      Lett. {\bf A127}, 112 (1988); P. B. Wiegmann, Phys. Rev. Lett.
	      {\bf 60}, 821 (1988); A. M. Polyakov, Mod. Phys. Lett. A {\bf 3},
	      325 (1988).

\bibitem{nohopf} T. Dombre and N. Read, Phys. Rev. B {\bf 38}, 7181
(1988); E. Fradkin and M. Stone, Phys. Rev B {\bf 38}, 7215
(1988); X. G. Wen and A. Zee, Phys. Rev. Lett. {\bf 61}, 1025 (1988).

\bibitem{sachread} N. Read and S. Sachdev, Phys. Rev. B {\bf 42}, 4568
(1990).

\bibitem{gross1} D. J. Gross, Nucl. Phys. B {\bf 132}, 439 (1978).

\bibitem{polya2} A. M. Polyakov, Phys. Lett. B {\bf 59}, 79 (1975); E. 
  Br\'ezin and J. Zinn-Justin, Phys. Rev. Lett. {\bf
  36}, 691 (1976); E. Br\'ezin and J. Zinn-Justin, Phys. Rev. B {\bf 14},
  3110 (1976).

\bibitem{halp} S. Chakravarty, B. I. Halperin and D. R. Nelson, Phys. Rev. B
               {\bf 39}, 2344 (1989); A. V. Chubukov, S. Sachdev and J. Ye, 
               Phys. Rev. B {\bf 49}, 11919 (1994).

\bibitem{gold} J. Goldstone, Nuovo Cimento {\bf 19}, 154 (1961);
Y. Nambu and G. Jona-Lasinio, Phys. Rev. {\bf 122}, 345 (1961);
J. Goldstone, A. Salam, and S. Weinberg Phys. Rev. {\bf 127}, 965
(1962).

\bibitem{bob1} R. B. Laughlin, Adv. Phys. {\bf 47}, 943 (1998);
B. A. Bernevig, D. Giuliano and R. B. Laughlin, An. of Phys. {\bf
311}, 182 (2004)

\bibitem{sachdev2} T. Senthil, A. Vishwanath, L. Balents, S. Sachdev and
                   M. P. A. Fisher, Science {\bf 303}, 1490 (2004).



\end{thebibliography}
\end{document}